\documentclass{article}

\usepackage[english]{babel}

\usepackage[letterpaper,top=2cm,bottom=2cm,left=3cm,right=3cm,marginparwidth=1.75cm]{geometry}

\usepackage{multicol}
\usepackage{multirow}
\usepackage{amssymb}
\usepackage{makecell}
\usepackage{tcolorbox}
\tcbuselibrary{skins}
\usepackage{graphicx}
\usepackage[table]{xcolor}
\usepackage{subcaption}
\usepackage{xcolor}
\usepackage{xspace}
\usepackage{wrapfig}
\usepackage{enumitem}
\usepackage{pifont}
\usepackage{colortbl}
\usepackage{tikz}
\usepackage{pgfplots}
\pgfplotsset{compat=1.18}
\usepackage{booktabs}
\usepackage{amsmath}
\definecolor{myblue}{HTML}{91ACE0}
\definecolor{mygreen}{HTML}{F5B7BF}
\definecolor{lightergray}{gray}{0.9}  

\usepackage[colorlinks=true, allcolors=blue]{hyperref}

\title{Who is Introducing the Failure? Automatically Attributing Failures of Multi-Agent Systems via Spectrum Analysis}

\newcommand{\tech}{\textsc{Famas}\xspace}

\begin{document}

\author{
Yu Ge\textsuperscript{1}\thanks{Both contributed equally to this work.}, 
Linna Xie\textsuperscript{1}\footnotemark[1], 
Zhong Li\textsuperscript{1}\thanks{Corresponding author.}, 
Yu Pei\textsuperscript{2}, 
Tian Zhang\textsuperscript{1}\thanks{Correspondence emails for all authors: yuge@smail.nju.edu.cn, xieln@smail.nju.edu.cn, lizhong@nju.edu.cn, yupei@polyu.edu.hk, ztluck@nju.edu.cn.} \\
\textsuperscript{1} Nanjing University
\textsuperscript{2} The Hong Kong Polytechnic University
}

\newcommand{\keywords}[1]{\textbf{Keywords:} #1}
\maketitle

\begin{abstract}
Large Language Model Powered Multi-Agent Systems (MASs) are increasingly employed to automate complex real-world problems, such as programming and scientific discovery. 
Despite their promising, MASs are not without their flaws. 
However, failure attribution in MASs—pinpointing the specific agent actions responsible for failures—remains underexplored and labor-intensive, posing significant challenges for debugging and system improvement.
To bridge this gap, we propose \tech, the first spectrum-based failure attribution approach for MASs, which operates through systematic trajectory replay and abstraction, followed by spectrum analysis.
The core idea of \tech is to estimate, from variations across repeated MAS executions, the likelihood that each agent action is responsible for the failure.
In particular, we propose a novel suspiciousness formula tailored to MASs, which integrates two key factor groups, namely the \textit{agent behavior group} and the \textit{action behavior group}, to account for the agent activation patterns and the action activation patterns within the execution trajectories of MASs. 
Through expensive evaluations against 12 baselines on the Who\&When benchmark, \tech demonstrates superior performance by outperforming all the methods in comparison.
\end{abstract}

\section{Introduction}
Large Language Model (LLM)-powered Multi-Agent Systems (MASs) are emerging as a novel software paradigm and are increasingly influential across diverse domains, such as software engineering~\cite{chatdev, openhands, swe-gpt}, scientific discovery~\cite{sciagent}, and general-purpose personal assistants~\cite{openmanus, fourney2024magentic-one}. 
Despite their promise, MASs are not without their flaws~\cite{mas_chanllenes_problems,microsoft_interactive_debugging_and_steering_MAS}. 
Particularly, recent studies have shown that MASs are susceptible to diverse failures, particularly in realistic, temporally evolving production environments~\cite{mast,trail}.
Therefore, effectively debugging such failures is essential to generate actionable insights for system refinement and reliability enhancement.

\noindent\textbf{Context.}
In this paper, we focus on failure attribution —the first phase of debugging~\cite{are_debugging_help_programmers}—as a critical step toward addressing this pressing need. 
The goal of failure attribution is to identify which action produced the system state that directly led to task failure~\cite{who&when}. 
Accurate failure attribution enables rapid identification of root causes, facilitating more effective debugging and system improvement. 
While MASs typically generate detailed logs documenting their operational processes~\cite{mast, who&when, agenttracer}, which provides a promising foundation for attribution, accurately interpreting these logs to attribute failures is difficult.
This difficulty are mainly two folds.
First, the problem-solving process in MASs often involves complex interactions among multiple LLM-powered agents, between agents and external tools, and within the internal reasoning processes of the LLMs themselves~\cite{trail}.
These interactions complicate system logs, challenging the interpretation of system behavior and hindering rapid root-cause identification. 
Second, the system actions and their resulting states are recorded in natural language within the log.
The inherent ambiguity of natural language further impedes precise characterization of operations and states.

\noindent\textbf{State of the art.}
Several studies have introduced fine-grained benchmarks to support failure attribution in MASs. 
For example, DevAI~\cite{devai} presents a coding benchmark structured around hierarchical user requirement, enabling identification of specific unmet requirements and offering a more nuanced evaluation compared to benchmarks that rely solely on final task success rate (e.g., SWE-Bench~\cite{swe-bench}).
However, the problem with these benchmarks is that they still merely provide additional metrics as reference points, while the process of failure attribution based on benchmark results remains a manual task.
More recent work has proposed employing the LLM-as-a-judge paradigm to diagnose MAS failures~\cite{mast,trail, who&when, agenttracer}. 
Despite these advancements, current LLM-based failure attribution methods achieve only limited success. 
For instance, the approach by Zhang et al.~\cite{who&when} attains an action-level failure attribution accuracy of less than 10\%.
This underscores the urgent need for more effective automated failure attribution methods in MASs.

\noindent\textbf{Our Approach.}
In this paper, we observe that the failure-responsible action and its resulting states frequently recur across repeated executions of the failed task. 
We hence propose a spectrum-based failure attribution approach for MASs, called \tech. Our approach is inspired by spectrum-based fault localization (SBFL)~\cite{sbfl_survey}, which is one of the prevalent fault localization technique in traditional software engineering~\cite{sbfl_popular}.
In SBFL, code entities executed more frequently by failing test cases are assigned higher spectrum scores, indicating a greater likelihood of being faulty. 
Analogously, given a failed execution trajectory of a task, \tech re-executes the task multiple times to collect a set of execution trajectories. 
It then computes spectrum scores for each action in the original failed trajectory by analyzing their occurrence frequency across these counterpart trajectories. 
Specifically, if an action appears more frequently in the counterpart trajectories, it receives a higher spectrum score, suggesting it is more likely to result in a erroneous system state that directly lead to task failure.

In \tech, there are two major challenges. 
The first one is how to accurately characterize the execution trajectories from extensive and verbose system logs.
To overcome this challenge, \tech introduces an LLM-based hierarchical clustering approach which segments each of the system logs into smal and manageable chunks, employs an LLM to analyze each chunk independently, and subsequently clusters the LLM outputs into a coherent sequence of agent–action–state triples that represent the execution trajectory. 
However, the execution trajectories of MASs typically exhibit greater heterogeneity and complexity than those of traditional programs, and thus the second challenge is how to account for these instinctive execution patterns in MASs to achieve accurate spectrum estimation. 
To overcome this challenge, we design a novel suspiciousness formula tailored to MASs, containing two key group metrics: the agent behavior group that captures the agent activation patterns and the action behavior group that captures the action activation patterns.

\noindent\textbf{Results.}
We verify the effectiveness of \tech based on the Who\&When benchmark~\cite{who&when} which consists of 184 failure traces from 127 MASs. 
We show that \tech achieves top performance on the Who\&When benchmark.
Specifically, \tech obtains a failure attribution of 29.35\% at the action-level, which are 49.13\% higher than the state-of-the-art technique by Zhang et al.~\cite{who&when}.
In addition, we extensively analyze the contribution of each design choices of \tech. 

\noindent\textbf{Summary.}
The main contribution of this paper are as follow:
 
\noindent\(\bullet\)\textbf{Approach.} We propose \tech, the first spectrum-based failure attribution approach that effectively identifies the root causes of failure execution trajectories in MASs. 

\noindent\(\bullet\)\textbf{Evaluation.} We demonstrate the effectiveness of \tech through comprehensive evaluations on the Who\&When benchmark, showing remarkable improvements over existing state-of-the-art failure attribution technique.

\noindent\(\bullet\)\textbf{Artifact.} We implement \tech into a tool with the same name and make it publicly downloadable to facilitate its easy application.

\section{Background}

\subsection{Large Language Model Powered Multi-Agent Systems}

In this work, we focus on the widely-adopted turn-based multi-agent protocol~\cite{metagpt, CAMEL, autogen}.
Specifically, let \(\mathcal{M}\) denote a Large Language Model (LLM)-powered multi-agent system (MAS), which consists of \(N\) agents  indexed by \(\mathcal{I} = \{1,2,\dots, N\}\).
These \(N\) agents operate in discrete time under the turn-based protocol, where exactly one agent performs an action at each time step.
Then, the MAS can be formally described as:
\[
\mathcal{M} = <\mathcal{I},\mathcal{S}, \mathcal{A}, \Psi,\phi>
\]
Here, \(\mathcal{S}\) denotes the set of possible of the system; \(\mathcal{A} = \mathcal{A}_1\cup \mathcal{A}_2\cup \dots \cup \mathcal{A}_n\) is the overall action space of the system, where \(\mathcal{A}_i\) is the set of actions specific to agent \(agent_i, i \in \mathcal{I}\).
\(\phi(t)\) is a function that returns the active agent at time step \(t\), thus specifying the turn-based schedule.
This active agent \(\phi(t)\) then selects an action \(a_t\in{\mathcal{A}_{\phi(t)}}\) conditioned on the current state \(s_t\).
Accordingly, the state-transition probability can de modeled as \(\Psi(s_{t+1}|s_t, a_t,\phi(t))\).

\noindent\textbf{MAS Execution Trajectory.} 
Consistent with existing literature~\cite{who&when, agenttracer}, we define the full execution trajectory of the MAS \(\mathcal{M}\) for completing a query \(\mathcal{Q}\) as \(\tau=(s_0, a_1,s_1,a_2,s_2,\dots,a_T,s_T )\), where \(T\) is a terminal time step or when the system  enters a terminating state, e.g., reaching the max interaction number. 
Furthermore, we employ a binary evaluation function \(\Omega(\tau) \in \{0,1\}\) to denote the result of the trajectory \(\tau\), where \(\Omega(\tau) = 1\) if the MAS successfully fulfills the query \(\mathcal{Q}\), and \(\Omega(\tau) = 0\) otherwise.

\subsection{Failure Attribution in MAS} 
Failure attribution in MAS is the process of identifying the components, such as a specific agent or a particular action, that directly lead to a task failure.
This is a crucial step for guiding systematic improvements, as it serves as the foundation for debugging and system refinement.
More specifically, given a failed trajectory \(\tau\) with \(\Omega(\tau)=0\), 
we define \textit{Failure Attribution in MAS} as the task of identifying the \textbf{decisive error} that constitutes the root cause of the failed trajectory \(\tau\).
A \textit{decisive error} is a specific action \(a_e\) that causes the system to enter an \textbf{error decisive state} \(s_e\);
Once the system enters the state \(s_e\)consequently, all subsequent actions conditioned on \(s_e\) diverge from the normal execution trajectory, inevitably leading to task failure.

 \begin{figure}[]
    \includegraphics[width=0.9\linewidth]{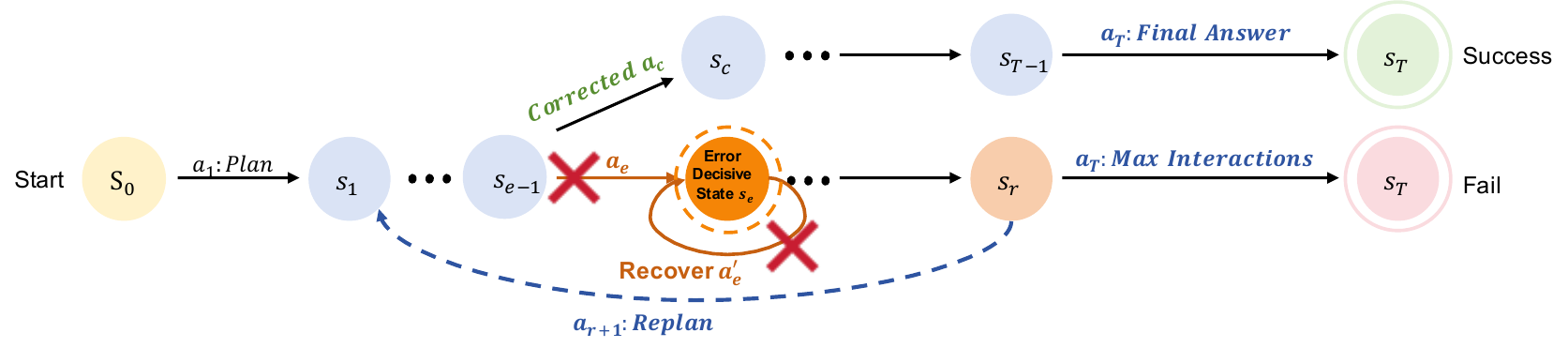}
    \caption{Fault Attribution in MAS.}
    \label{fig:state_transition}
\end{figure}

We analyze the root cause of failed trajectories from a state perspective for the following reasons.
In MASs, the same action type (e.g., semantically equivalent actions) may appear multiple times in a trajectory due to the self-improvement mechanisms that enable recovery and re-planning during execution failures.
However, if these actions fails to bypass the error decisive state, downstream actions remains affected by this state, preventing task completion.
For better illustration, Figure~\ref{fig:state_transition} depicts successful and failed trajectories for a query. 
As shown, the system recovery occurs only wehn a correct action \(a_c\) enables a transition from the error-deciding state to a correct state; otherwise, if the action \(a_e\) (or \(a'_e\)) that drives the system into the error decisive state \(s_e\) remains in the trajectory, subsequent actions continue to be affected, inevitably leading to task failure.
In this work, we call the action \(a_e\) that triggers the transition into \(s_e\) as \textbf{error decisive action}, denoted by "\(\xrightarrow{a_e} s_e\)". 
In addition, we refer the agent that produces the error decisive action to failure-responsible agent.
Therefore, by attributing the error decisive state and its corresponding error decisive actions, one can better understanding why the system fails and which actions cause the failure, facilitating more effective debugging and system improvement.

In practice, the trajectory $\tau$ is typically represented as an execution log~\cite{mast, who&when, agenttracer}---a complete record of the conversation or interaction history among agents during their attempt to solve query $\mathcal{Q}$. These logs serve as the primary evidential basis for failure analysis and attribution. Formally, given a query $\mathcal{Q}$, we define the failure execution log $\mathcal{L}_{\tau}$ as an alternative representation of $\tau$:
\begin{equation}
    \mathcal{L}_{\tau} = (s_0, \eta_1,\eta_2,\dots, \eta_T), \quad \text{where } \eta_t = <\phi(t), a_t, s_t>.
    \label{eq:trajectory}
\end{equation}
Here, $\phi(t)$ denotes the agent active at interaction step $t$, and $s_0$ represents the initial state given the query $\mathcal{Q}$. Each subsequent element $\eta_t = \langle \phi(t), a_t, s_t\rangle$ in $\mathcal{L}_{\tau}$ indicates that agent $\phi(t)$ performed action $a_t$, resulting in a transition of the MAS to state $s_t$ at interaction step $t$. Accordingly, the task of \textit{failure attribution} in MAS is to identify the specific agent–action–state tuple $<\phi(e), a_e, s_e>$ in $\mathcal{L}_{\tau}$ that constitutes the \textit{decisive error}. 

\subsection{Spectrum-Based Fault Localization}~\label{sec:sbfl}
\begin{wraptable}{r}{0.4\textwidth}
    \centering
    \scriptsize
    \caption{Representative SBFL formulas.}
    \label{tab:sbfl-formulas}
    \begin{tabular}{l l}
    \toprule
    \textbf{Formula} & \textbf{Suspiciousness Score $S(c)$} \\
    \midrule
    Ochiai & $\frac{n_{cf}}{\sqrt{(n_{cf}+n_{u_f})*(n_{cf}+n_{cs})}}$ \\[6pt]
    Tarantula    & $\frac{n_{cf}}{n_{cf}+n_{u_f}}/(\frac{n_{cf}}{n_{cf}+n_{u_f}} + \frac{n_{cs}}{n_{cs}+n_{u_s}})$ \\[6pt]
    Jaccard   & $\frac{n_{cf}}{n_{cf}+n_{u_f}+n_{cs}}$ \\[6pt]
    Dstar2   & $\frac{n^2_{cf}}{n_{cs}+n_{u_f}}$ \\[6pt]
    Kulczynski2 & $\frac{1}{2}*(\frac{n_{cf}}{n_{cf}+n_{u_f}}+\frac{n_{cf}}{n_{cf}+n_{cs}})$\\
    \bottomrule
    \end{tabular}
\end{wraptable}
In this work, we propose a novel specturm-based failure attribution approach \tech for MASs, inspired by the traditional Spectrum-Based Fault Localization (SBFL)~\cite{sbfl_survey}.
To contextualize our approach, we briefly review SBFL here.
SBFL is a widely adopted and lightweight debugging technique primarily used in software diagnosis to identify faults in programs. The core idea of SBFL is to discover statistical correlations between system failures and the activity of different parts of the system. By running tests and recording test outcomes along with coverage information, SBFL can provide developers with a ranked list of potentially faulty components by their suspiciousness. 

More specifically, given a set of program components \(C\) under analysis, an SBFL technique executes a set of tests \(T\) and records execution outcomes in: 1) a coverage matrix \(M\in\{0,1\}^{|T|\times|C|}\) where \(m_{ij}=1\) if the component \(c_j\in{C}\) is executed by the test \(t_i\in{T}\) and 2) an error vector \(E\in\{0,1\}^{|T|}\) where \(e_i = 0\) if the test case \(t_i\) fails. 
Then, it derives the following basic statistical metrics from the coverage matrix and error vector: 1) $n_{cf}$: The number of failed test cases that covered the component; 2) $n_{uf}$: The number of failed test cases that uncovered the component; 3) $n_{cs}$: The number of successful test cases that covered the component; and 4) $n_{cf}$: The number of successful test cases that uncovered the component.
Based on these four metrics, the SBFL technique employs a suspiciousness formula to compute a suspiciousness score \(S(c)\) for each component. 
Finally, the components are ranked by \(S(c)\) to generate fault localization results.
Commonly used formulas include Ochiai\cite{Ochiai}, Tarantula \cite{tarantula}, Jaccard \cite{Jaccard}, Dstar2 \cite{Dstar2} and Kulczynski2 \cite{Kulczynski2}, summarized in Table~\ref{tab:sbfl-formulas}.

\section{Methodology}
In this section, we use an example to illustrate the motivation and design philosophy of \tech.

\begin{figure}[htbp]
    \centering
    \begin{subfigure}{\linewidth}
        \centering
        \includegraphics[width=\linewidth]{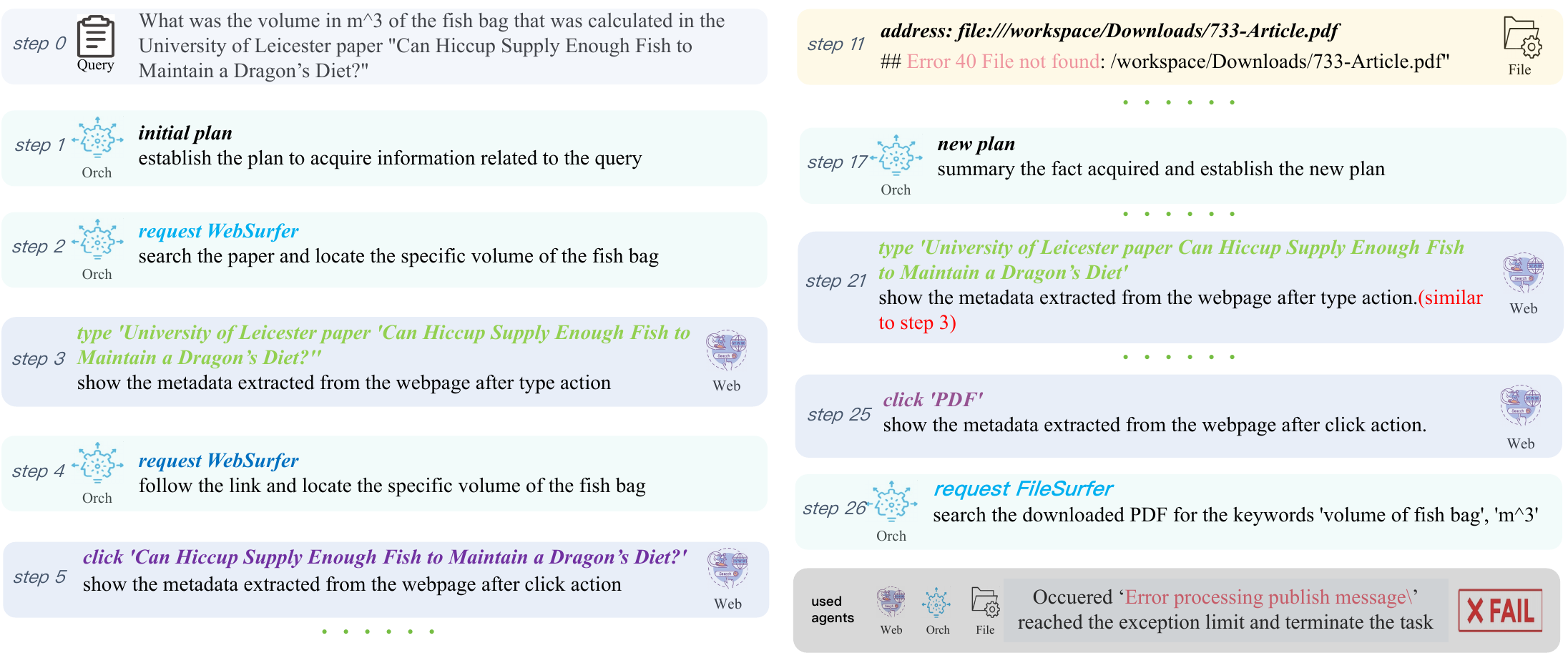}
        \caption{A failed execution trajectory where multiple agents collaborate but the task does not reach the correct outcome.}
        \label{fig:fail-execution-trajectory}
    \end{subfigure}

    \begin{subfigure}{\linewidth}
        \centering
        \includegraphics[width=0.8\linewidth]{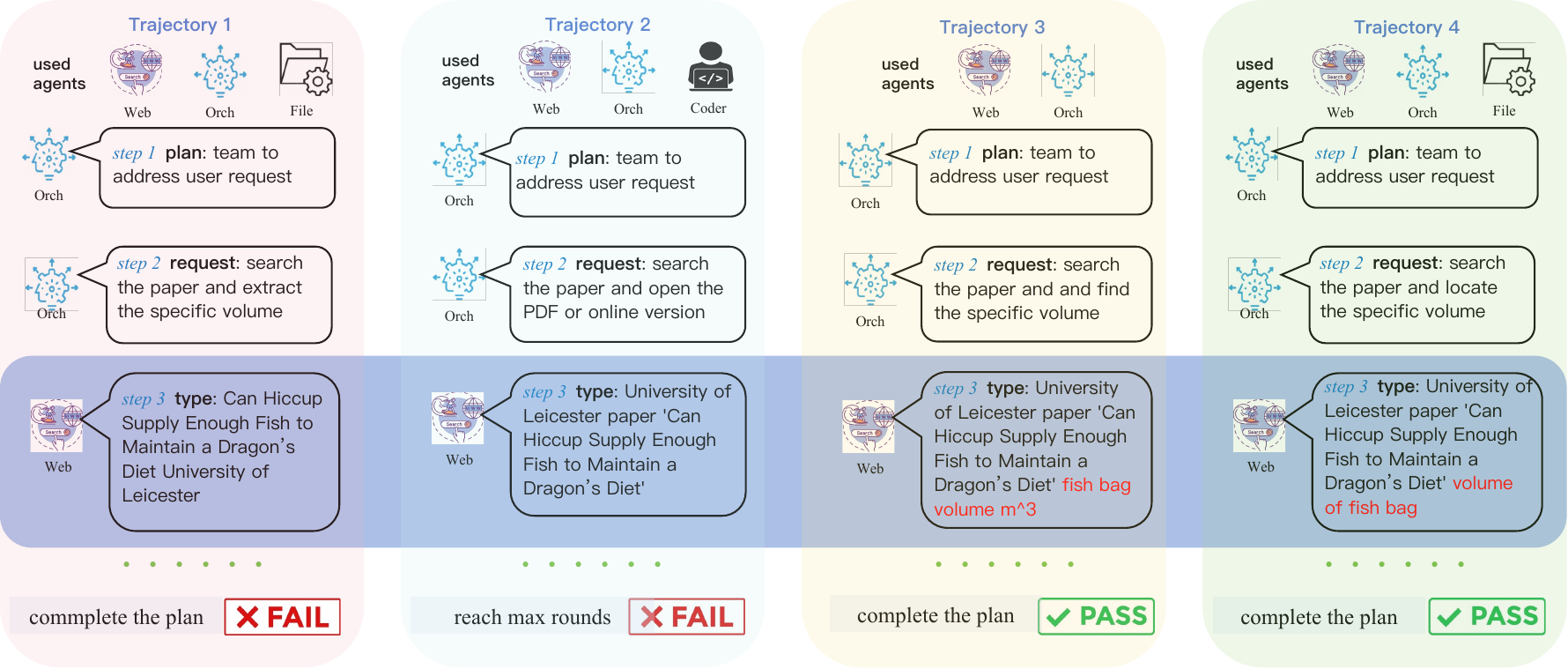}
        \caption{Four additional runs of the same task, including two failures and two successes.}
        \label{fig:extra-trajectories}
    \end{subfigure}

    \caption{Execution trajectories of a real-world task from the GAIA dataset by the MAS \textit{MagenticOne}.}
    \label{fig:motivation-example}
\end{figure}

\noindent\textbf{Example}. 
Figure~\ref{fig:fail-execution-trajectory} presents a simplified log illustrating a failure case when a MAS fails to execute a query task. 
Specifically, the MAS is assigned to answer the question ``What was the volume in \(m^3\) of the fish bag that was calculated in the university of Leicester paper `Can Hiccup Supply Enough Fish to Maintain a Dragon’s Diet?'''. 
However, the MAS improperly performs the web search action (Step 3) using an inaccurate search description, resulting in erroneous search results that subsequently misdirected downstream actions. 
Although the MAS incorporates a self-improvement mechanism that is able to retry the web search operation (Step 21), it still struggle to provide accurate search queries.
Consequently, the system persistently returns incorrect search results, ultimately failing to complete the task.
Failure attribution aims to identify the state of erroneous search results and the corresponding search actions that lead to this state, enabling the refinement of actions to optimize system execution.

\noindent\textbf{Manual Failure Attribution}. Analyzing the system log manually to attribute the failures presents a significant challenge. 
As illustrated in Figure \ref{fig:fail-execution-trajectory}, the log is extensive and verbose, comprising approximately 16k tokens that include not only core actions from the system execution process but also extraneous content such as non-operational entries, metadata, and auxiliary system outputs. 
This complexity impedes the precise identification of relevant action sequences and their corresponding result states. 
In addition, accurate failure attribution requires expert-level knowledge about the architecture and behavior of the system. 
Therefore, manual failure attribution is a highly time-consuming and expertise-intensive process, rendering it impractical in real-world applications.

\noindent\textbf{LLM-based Failure Attribution}. 
Recently, Zhang et al.~\cite{who&when} have explored leveraging the LLM-as-a-judge paradigm to analyze system logs for failure attribution. 
More specifically, it employs three carefully designed prompting strategies to instruct a LLM to generate failure attribution results from the logs. 
However, the LLM struggles to accurately identify error decisive state and action due to the extensive and often noisy context present in lengthy MAS logs~\cite{llm_fail_on_long_context_1, llm_fail_on_long_context_2}. 
Compounding this issue, the logs frequently contain misleading entries that can mislead the analysis of the LLM. 
For instance, the approach by Zhang et al. incorrectly attribute failure to a file-reading action and its associated state (Line 11), where superficially salient terms such as “Error” and “file not found” appear, causing the LLM to overlook the true root cause—potentially embedded in a seemingly innocuous entry (e.g., Lines 3 and 21). 
Notably, the apporach by Zhang et al.~\cite{who&when} only obtains an action-level failure attribution accuracy of 7.02\% for handcrafted MASs, which is only marginally better than a random baseline of 4.16\%. 
More detailed discussion can be found in Section~\ref{sec:rq2}.

\noindent\textbf{Our Idea.}
Our key observation is that the error decisive state and action in a failed trajectory also frequently recur across repeated executions of the task.
Consider the failure case in Figure \ref{fig:fail-execution-trajectory}, we repeatedly re-execute the task and collect execution trajectories, with Figure \ref{fig:extra-trajectories} displaying simplified logs from these runs. 
From Figure \ref{fig:extra-trajectories}, we can observe that failed runs frequently include a specific web search action that returns erroneous results, whereas successful trajectories typically invoke an alternative, more effective search. 
Therefore, it is intuitive to attribute the failure trajectory in Figure \ref{fig:fail-execution-trajectory} through evaluating the frequency of actions and their resulting states across the aggregated execution trajectories.

This observation is analogous to traditional spectrum-based fault localization (SBFL)~\cite{sbfl_survey}. 
In particular, SBFL assumes that code entities executed more frequently by failing tests are more likely to be faulty. 
It then computes a suspiciousness core for each code entity using aggregated test execution data to guide fault localization. 
Therefore, \textit{our overarching idea to identify the error decisive state and action of a specific failed trajectory is to conduct spectrum analysis on multiple trajectories collected through repeated execution of the corresponding task}. 
To realize the idea, it is important to address the following two challenges.

\noindent\(\bullet\)\textbf{C1: How to accurately characterize the execution trajectories from system logs}?
As discussed earlier, the actions and their resulting states are specified using natural language within execution logs. 
However, the flexibility of natural language indicates that the semantic-equal actions and their resulting states can be described in different ways. 
For instance, the Step 3 of trajectory 1 and 2 in Figure \ref{fig:extra-trajectories} can be slightly different to the step 3 in Figure \ref{fig:fail-execution-trajectory} with respect to punctuation and word order, while they present the same semantic.
Consequently, such variations make it difficult to consistently extract agent–action–state triples \(\langle agent_i, a, s \rangle\) from logs, introducing noisy signals for specturm analysis.

To address this challenge, we leverage the power text analysis capabilities of LLMs~\cite{llm_power_information_extraction} to transform system logs into execution trajectories.
However, the LLMs face difficulties in accurately extracting entities when processing lengthy inputs~\cite{llm_fail_on_long_context_1, llm_fail_on_long_context_2}. 
Therefore, \tech applies an LLM separately to each system log and splits these logs into manageable chunks to extract primitive agent-action-state triples. 
In addition, we introduce a hierarchical clustering approach to refine these primitive triples by first identifying distinct agents and then categorizing different action-state pairs. 
Such a clustering mechanism further helps eliminate variations among multiple LLM outputs, yielding consistent and structured trajectories suitable for downstream spectrum-based fault analysis.

\noindent\(\bullet\)\textbf{C2: How to accurately estimate the spectrum scores?}
Compared to traditional programs, MASs typically exhibit more complex and diverse execution trajectories, involving interactions among multiple agents, tool invocations, and the agents’ internal reasoning processes~\cite{trail}. 
Intuitively, it is sub-optimal to directly apply the existing SBFL techniques~\cite{ Ochiai, tarantula, Jaccard, Dstar2, Kulczynski2} for failure attribution in MASs. 
Experimental validation can be found in Section~\ref{sec:rq2}.

To address this challenge, we introduce two groups of metrics specifically designed for MAS environments: the Agent Behavior Group and the Action Behavior Group. 
These metrics capture distinctive aspects of MAS failures that traditional SBFL techniques overlook. 

The Agent Behavior Group metrics are designed to address the fundamental challenge of the heterogeneity in MASs, wherein an agent is first activated and subsequently selects an action.
Our observation is that different agnets often exhibit vastly different activity levels, direct frequency-based comparisons would inherently bias results towards more active agents.
For examples, in Figure~\ref{fig:motivation-example}, the FileSurfer agent activates only in the trajectory of Figure~\ref{fig:fail-execution-trajectory} while remains deactivated across all repeated executions in Figure~\ref{fig:extra-trajectories}, resulting it with a very low coverage ratio and thus making it be easily ignored in the spectrum analysis.
Therefore, we propose two complementary metrics to ensure fair and meaningful comparison across diverse agent types.
First, Agent-Action Coverage Ratio ($\gamma$) assesses how widely distributed an action is across different execution contexts involving the agent, distinguishing between consistently used functions and situation-specific behaviors. 
Second, The Agent-Action Frequency Proportion ($\beta$) measures how frequently a specific action appears relative to all actions performed by that agent, normalizing for variations in agent activity levels and identifying which actions are core to an agent's behavior.


The Action Behavior Group metrics are designed to capture action characteristics in MASs.
Our observation is that the same type of actions can repeatedly occur within an execution trajectory due to the self-improvement mechanisms of MASs. 
For example, as shown in Figure~\ref{fig:fail-execution-trajectory}, step 21 re-executes an action of the same type as step 3 following a re-planning phase. 
Both actions lead to similar system states that misdirect the system.
Therefore, we propose the Local Frequency Enhancement Factor ($\alpha$), which specifically amplifies suspicious actions that appear unusually frequently within individual failing trajectories, to consider the action repeatability.
However, certain meta-actions, such as planning actions in Figure~\ref{fig:motivation-example}, play foundational roles in initiating execution and are therefore present in both successful and failing trajectories.
To account for such globally recurrent actions, we further propose the $\lambda$-Decay SBFL Coefficient, which captures global frequency patterns across multiple executions by applying exponential decay to repeated occurrences, preserving strong signals while attenuating redundant repetitions.

	\begin{figure*}[!tb]
		\centering
		\includegraphics[width=\linewidth]{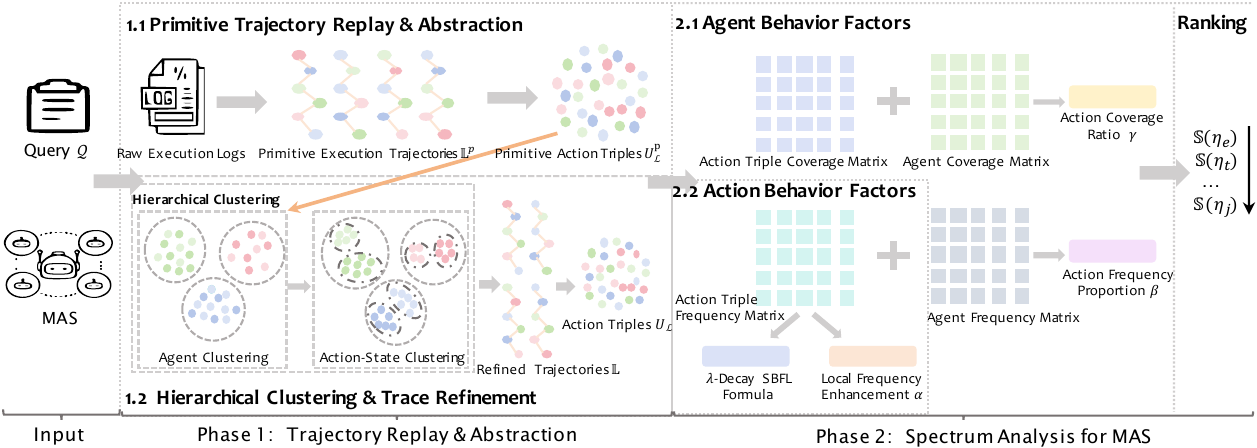}
		\caption{Overview of \tech.}
		\label{fig:overview}
	\end{figure*}
\section{Design}

Figure~\ref{fig:overview} illustrates the workflow of \tech. 
Specifically, \tech comprises two main technical modules/phases: \textit{Trajectory Reply} \& \textit{Abstraction} and \textit{Spectrum Analysis}. 
Given a task query \(\mathcal{Q}\) on which the MAS \(\mathcal{M}\) fails, producing an execution trajectory \(\tau\), \tech identifies the decisive error within \(\tau\) as follows. 
In the \textit{Trajectory Replay} \& \textit{Abstraction}, \tech first re-executes the query task \(\mathcal{Q}\) multiple times to collect raw execution logs from repeated runs.  
These natural language logs are then processed by an LLM-based hierarchical clustering procedure, which transforms them into a structured set of execution trajectories, denoted as \(\mathbb{L}\).
This process mitigates the inherent ambiguity of natural language, yielding more precise trajectories for the following specturm analysis.
Then in the \textit{Spectrum Analysis}, \tech conducts spectrum analysis on the set of execution trajectories \(\mathbb{T}\) to estimate the spectrum scores \(\mathbb{S}\) of each actions.
Particularly, we introduce a novel suspiciousness formula that evaluates actions at both the agent level and the action level. 
Finally, \tech ranks entities in \(\tau\) by their spectrum scores \(\mathbb{S}\) and identifies the top-ranked entity as the  attribution result.

\subsection{Phase 1: Trajectory Replay \& Abstraction}

This phase is responsible for acquiring execution logs from multiple runs of the MAS \(\mathcal{M}\) and transforming them into concrete and  structured execution trajectories suitable for spectrum-based fault localization. 
Specifically, the process contains two main steps: \textit{primitive trajectory replay} \& \textit{abstraction} and \textit{hierarchical clustering} \& \textit{trajectory refinement}.

\subsubsection*{Step 1.1: Primitive Trajectory Replay \& Abstraction}
Given a query $\mathcal{Q}$ with a failure log $l_0$, we replay the trajectories of $k$ independent runs to collect the corresponding execution logs. This yields a raw execution log suite
\[
L = \{ l_0, l_1, l_2, \dots, l_k \},
\]
where $l_0$ corresponds to the failing execution and $\{ l_1, \dots, l_k \}$ are logs from subsequent runs, which may succeed or fail depending on nondeterministic agent interactions. Collectively, this log suite captures a spectrum of system behaviors that provides the necessary variability for fault localization.

The raw logs in $L$ are unstructured and heterogeneous, and thus not directly amenable to analysis. We therefore transform each log $l_i$ into a structured trajectory $\mathcal{L}_{\tau_i}$ (see Equation~\ref{eq:trajectory}). To achieve this, we employ an LLM to perform semantic parsing, converting each record in the log into a canonical triple $\langle \texttt{AGENT}, \texttt{ACTION}, \texttt{STATE} \rangle$, where \texttt{AGENT} denotes the entity that initiates an active agent, \texttt{ACTION} specifies the concrete behavior executed by the agent, and \texttt{STATE} represents the resulting system state after the action. For example, if the web agent performs the action \texttt{search}, the resulting \texttt{STATE} could be a screenshot of the webpage retrieved by the MAS. 
More specifically, we split each log \(l_i\in{L}\) to small and manageable chunks and prompt an LLM to processes these chunks sequentially to yield candidate triples. Formally, let 
\[
\mathbb{L}^p = \{ \mathcal{L}_{\tau_0}^p, \mathcal{L}_{\tau_1}^p, \dots, \mathcal{L}_{\tau_k}^p \}, \quad 
\mathcal{L}_{\tau_i}^p = \{s_0, \eta_1^p, \dots, \eta_T^p\}, \quad 
\eta_j^p = \langle \phi(j), a_j^p, s_j^p \rangle
\]
denote the collection of primitive \textit{execution trajectories} extracted from all runs, where $\phi(j)$ is the acting agent, $a_j^p$ the executed action, and $s_j^p$ the resulting state.

\subsubsection*{Step 1.2: Hierarchical Clustering \& Trajectory Refinement.}
After the primitive abstraction of each raw execution log, we further employ a hierarchical clustering approach to refine the candidate triples in \(\mathbb{L}^p\).
This is motivated by the observation that semantically equivalent triples may still appear in different surface forms due to the variations of LLM outputs. 
Specifically, we aggregate all primitive agent–action–state triples $\eta^p$ obtained from the \textit{abstracted primitive trajectories $\mathbb{L}^p$}:
\[
U^p_\mathbb{L} = \bigcup_{i=0}^k \mathcal{L}_{\tau_i}^p = \{\langle i, a, s \rangle|\langle i, a, s \rangle \in \mathcal{L}^p_{\tau_j}, j\in[0,k]\}.
\] 
Based on $\mathbb{L}^p$, the clustering approach first groups the triples by their agent identifier (i.e., the index of the agent that initiates the action).
Then, it further groups the triples within each agent group by prompting an LLM to analyze semantic similarity of their action–state descriptions, producing consistent abstractions for spectrum analysis.

\subsubsection*{Trajectory Output.}
Once clustering is complete, each primitive trajectory \(\mathcal{L}_{\tau_i}^p\) is \textit{refined} into a final abstracted trajectory \(\mathcal{L}_{\tau_i}\) by replacing each primitive triple with its corresponding cluster representative. This refinement eliminates redundancy, consolidates semantically equivalent behaviors, and yields a concrete, structured \textit{behavioral spectrum} suitable for subsequent spectrum analysis.
Formally, the final set of abstracted execution trajectories is:
\[
\mathbb{L} = \{ \mathcal{L}_{\tau_0}, \mathcal{L}_{\tau_1}, \dots, \mathcal{L}_{\tau_k} \}, \quad \mathcal{L}_{\tau_i} = \{s_0, \eta_0, \eta_1,\dots,\eta_T\},\quad \eta_j = <\phi(j), a_j, s_j>
\] 
where each \(\mathcal{L}_{\tau_i}\) is a refined trajectory of representative agent–action–state triples that captures the essential behavior of the MAS. Furthermore, we output the universe of all unique agent-action-state triples across executions $U_\mathbb{L} = \bigcup_{i=0}^k \mathcal{L}_{\tau_i}$ for facilitating the subsequent spectrum analysis.

\subsection{Phase 2: Spectrum Analysis in MAS}
\label{sec:fa4MAS}
With the set of abstracted execution trajectories \(\mathbb{L}\), we then conduct spectrum analysis on these trajectories to attribute failures.
Specifically, we propose a novel suspiciousness formula that integrates four key metrics to capture their execution patterns.
This four metrics organized into two categories: the \textit{Agent Behavior Group}, which includes Action Coverage Ratio ($\gamma$) and Action Frequency Proportion ($\beta$); and the \emph{Action Behavior Group}, which includes the Global Frequency Decay ($\lambda$-Decay SBFL Formula) and the Local Frequency Enhancement ($\alpha$). 
Next, we elaborate on the suspiciousness formula as well as these four metrics.

\subsubsection{Matrices for Spectrum Analysis}
Prior to calculating the suspiciousness score for each agent-action-state triple in the universal set $U_{\mathbb{L}}$, we construct several matrices that collectively encode the execution spectrum of the MAS. These include:

\noindent\(\bullet\) A binary coverage matrix $\mathbf{C}_\eta \in \{0,1\}^{(k+1) \times m}$ where rows correspond to execution trajectories $\mathcal{T}$ ($k+1 = |\mathcal{T}_{\text{succ}}| + |\mathcal{T}_{\text{fail}}|$), columns correspond to unique agent-action-state triples in $U_{\mathbb{L}}$ ($m = |U_{\mathbb{L}}|$), and each element $c_{ij} = 1$ if agent-action-state $\eta_j \in U_{\mathbb{L}}$ appears in execution log $\mathcal{L}_{\tau_i} (\tau_i\in \mathcal{T})$, and $0$ otherwise. 

\noindent\(\bullet\)  A frequency matrix $\mathbf{F_\eta} \in \mathbb{N}^{(k+1) \times m}$ where each element $f_{ij}$ records the occurrence count of agent-action-state triple $\eta_j$ in trajectory $\mathcal{L}_{\tau_i}$.

\noindent\(\bullet\)  An outcome vector $\mathbf{O} \in \{0,1\}^{k+1}$ where $o_i = 1$ indicates trajectory $\tau_i$ succeeded ($\Omega(\tau_i)=1$) and $o_i = 0$ indicates trajectory $\tau_i$ failed ($\Omega(\tau_i)=0$).

Additionally, we construct analogous matrices at the agent level ($\mathbf{C}_{\text{agent}}$, $\mathbf{F}_{\text{agent}}$) where columns correspond to individual agents rather than specific agent-action-state triples, enabling analysis of agent-level behavioral patterns alongside the fine-grained agent-action-state triple analysis.
\begin{small}
$$
\mathbf{C_\eta} = \begin{bmatrix}
	& \eta_1 & \eta_2 & \cdots & \eta_m \\
	\tau_0 & c_{01} & c_{02} & \cdots & c_{0m} \\
	\tau_1 & c_{11} & c_{12} & \cdots & c_{1m} \\
	\vdots & \vdots & \vdots & \ddots & \vdots \\
	\tau_k & c_{k1} & c_{k2} & \cdots & c_{km}
\end{bmatrix},\quad\quad
\mathbf{F_\eta} = \begin{bmatrix}
	& \eta_1 & \eta_2 & \cdots & \eta_m \\
	\tau_0 & f_{01} & f_{02} & \cdots & f_{0m} \\
	\tau_1 & f_{11} & f_{12} & \cdots & f_{1m} \\
	\vdots & \vdots & \vdots & \ddots & \vdots \\
	\tau_k & f_{k1} & f_{k2} & \cdots & f_{km}
\end{bmatrix},
 \quad\quad
\mathbf{O} = \begin{bmatrix}
	\Omega(\tau_0) \\
	\Omega(\tau_1) \\
	\vdots \\
	\Omega(\tau_k)
\end{bmatrix}
$$$$
\mathbf{C_{agent}} = \begin{bmatrix}
	& agent_1 & agent_2 & \cdots & agent_n \\
	\tau_0 & c_{01} & c_{02} & \cdots & c_{0n} \\
	\tau_1 & c_{11} & c_{12} & \cdots & c_{1n} \\
	\vdots & \vdots & \vdots & \ddots & \vdots \\
	\tau_k & c_{n1} & c_{k2} & \cdots & c_{kn}
\end{bmatrix},\quad\quad
\mathbf{F_{agent}} = \begin{bmatrix}
	& agent_1 & agent_2 & \cdots & agent_n \\
	\tau_0 & f_{01} & f_{02} & \cdots & f_{0n} \\
	\tau_1 & f_{11} & f_{12} & \cdots & f_{1n} \\
	\vdots & \vdots & \vdots & \ddots & \vdots \\
	\tau_k & f_{k1} & f_{k2} & \cdots & f_{kn}
\end{bmatrix}.
$$
\end{small}
\subsubsection{Agent Behavior Group}
The Agent Behavior Group introduces two complementary metrics that address agent heterogeneity: For a given agent-action-state triple $\eta_j = \langle agent_i, a, s \rangle$ and its corresponding agent $i$:
\begin{equation}
    \gamma = nc_{\eta_j} / nc_{agent_i},
    \label{eq:acr}
\end{equation}
where $nc_{\eta_j} = \sum_{p=0}^{k} c_{pj}$ denotes the number of trajectories containing triple $\eta_j$ (with $c_{pj}$ being elements from the coverage matrix $\mathbf{C}_{\eta}$), and $nc_{agent_i} = \sum_{p=0}^{k} c_{pi}$ represents the number of trajectories where agent $agent_i$ appears (with $c_{pi}$ being elements from the agent-level coverage matrix $\mathbf{C}_{\text{agent}}$).

\begin{equation}
    \beta = f_{\eta_j} / f_{agent_i},
    \label{eq:afr}
\end{equation}
where $f_{\eta_j} = \sum_{p=0}^{k} f_{pj}$ denotes the global frequency of triple $\eta_j$ (with $f_{pj}$ being elements from the frequency matrix $\mathbf{F}_{\eta}$), and $f_{agent_i} = \sum_{p=0}^{k} f_{pi}$ represents the total frequency of all actions performed by agent $agent_i$ (with $f_{pi}$ being elements from the agent-level frequency matrix $\mathbf{C}_{\text{agent}}$).

The Action Coverage Ratio ($\gamma$) measures the prevalence of a specific action-state pair across an agent's executions, where high values indicate consistent behavior and low values suggest context-dependent operations. The Action Frequency Proportion ($\beta$) quantifies an action's relative importance within an agent's behavioral repertoire, with high values indicating core functionality and low values suggesting peripheral activities. These metrics operate synergistically—$\beta$ captures behavioral intensity while $\gamma$ reflects contextual breadth—enabling differentiation between concentrated core errors and widely distributed failures. This dual approach eliminates agent activity bias while providing nuanced insights into failure patterns characteristic of multi-agent systems.

\subsubsection{Action Behavior Group.} 
The Action Behavior Group introduces two complementary metrics that address action repeatability.
The Local Frequency Enhancement Factor ($\alpha$) amplifies intra-trajectory anomalies for a specific failure trajectory $\tau_i$:
\begin{equation}
\alpha_{\tau_i} = 1 + \log_{1/\lambda}\big(f_{ij}\big)
\label{eq:enhance}
\end{equation}
These metrics work synergistically: the $\lambda$-decay captures cross-trajectory frequency patterns that distinguish failure-correlated actions from ubiquitous background operations, while the $\alpha$-factor emphasizes actions that exhibit abnormal repetition within specific failing trajectories. 

Complementarily, The $\lambda$-Decay SBFL Coefficient incorporates global frequency sensitivity through exponential decay weighting ((the definition of SBFL refers to Section \ref{sec:sbfl}). For a given agent-action-state triple $\eta_j = \langle agent_i, a, s \rangle$:
\begin{equation}
n_{cf}^{\lambda} = \sum_{p=0}^k 
\begin{cases}
\lambda^{f_{pj}-1}, & \text{if} \ \  f_{pj} > 0  \text{ and }\textbf{O}(p) = 0,\\
0, & otherwise
\end{cases}
\quad
n_{cs}^{\lambda} = \sum_{p=0}^k 
\begin{cases}
\lambda^{f_{pj}-1}, & \text{if} \ \  f_{pj} > 0  \text{ and }\textbf{O}(p) = 1,\\
0, & otherwise
\end{cases}
\label{eq:decy}
\end{equation}
where $f_{pj}$ being elements from the frequency matrix $\mathbf{F}_\eta$ and $\lambda \in (0.5,1)$ is a decay factor that preserves initial occurrence signals while attenuating redundant repetitions. Substituting $n{cf}$ and $n_{cs}$ with $n_{cf}^{\lambda}$ and $n_{cs}^{\lambda}$ yields the $\lambda$-decay SBFL formulation.
This dual approach enables detection of both globally prevalent fault patterns and locally concentrated anomalies, addressing fundamental limitations of binary occurrence counting in multi-agent environments. 

\subsubsection{Suspiciousness Calculation and Ranking}
The suspiciousness score for each agent-action-state triple $\eta_j = \langle agent_i, a, s \rangle$ within the failed execution trajectory $\tau_0$ is computed by integrating metrics from both behavior groups using Kulczynski2 \cite{Kulczynski2} (see Table \ref{tab:sbfl-formulas}) as the base SBFL formula. The combined score $\mathbb{S}(\eta_j)$ is defined as:
\begin{equation}
\mathbb{S}(\eta_j) = \left[\alpha_{\tau_0}(\eta_j) \cdot \text{Kulczynski2}^{\lambda}(\eta_j)\right] \cdot \left[1+\beta(\eta_j)\right] \cdot\left[1+ \gamma(\eta_j)\right]
\label{final_eq}
\end{equation}
where the $\lambda$-decay enhanced Kulczynski2 metric is calculated as:
\begin{equation}
\text{Kulczynski2}^{\lambda}(\eta_j) = \frac{1}{2} \left( \frac{n_{cf}^{\lambda}(\eta_j)}{n_{cf}^{\lambda}(\eta_j) + n_{uf}} + \frac{n_{cf}^{\lambda}(\eta_j)}{n_{cf}^{\lambda}(\eta_j) + n_{cs}^{\lambda}(\eta_j)} \right)
\end{equation}
All agent-action-state triples are ranked in descending order based on their suspiciousness scores:
\begin{equation}
\mathbb{R} = \left\langle (\eta_j, \mathbb{S}(\eta_j)) \mid \eta_j \in U_{\mathbb{L}} \right\rangle_{\downarrow \mathbb{S}}
\end{equation}

The ranking prioritizes agent-action-state combinations that exhibit both strong statistical correlation with failures (captured by Kulczynski2$^{\lambda}$) and significant behavioral anomalies (captured by $\alpha$, $\beta$, and $\gamma$), providing a comprehensive fault localization approach tailored for multi-agent systems. Following strict evaluation criteria, only the top-1 ranked triple is considered as the final output for fault attribution.

\section{Experiments}
We evaluate \tech on the following research questions:

\noindent\(\bullet\) \textbf{RQ1}: How effective and efficient is \tech in failure attribution for MASs?

\noindent\(\bullet\) \textbf{RQ2:} How does the accuracy of our SBFL-based \tech compare with random approach, LLM-based approaches and other SBFL formulas in multi-agent system failure attribution?


\noindent\(\bullet\) \textbf{RQ3:} How do the parameters of \tech affects its effectiveness?


\subsection{Evaluation Setup}\label{sec:exp_setup}

\noindent\textbf{Benchmark.}
We evaluate \tech on the recently proposed Who\&When benchmark~\cite{who&when}.
This benchmark comprises 184 failure logs  from 127 MASs, including 126 algorithmically generated systems based on the AG2 framework~\cite{AG2_2024} and one hand-crafted system derived from the Magnetic-One platform~\cite{fourney2024magentic-one}.
This benchmark encompasses a wide range of realistic scenarios.
Furthermore, each log in this benchmark is carefully annotated by three human experts through a multi-round consensus procedure to identify the agents and actions responsible for the failure, ensuring high annotation reliability.
Please note that, to the best of our knowledge, the Who\&When benchmark is currently the only publicly available benchmark for failure attribution in MASs.

\noindent\textbf{Baselines.}
In our experiments to address RQ2, we consider in total 12 compared approaches, including one random approach, six LLM-based approaches and 5 variants of \tech based on representative SBFL formulas.

\noindent\(\bullet\) \textit{Random Failure Attribution.}
This refers to the most basic baseline, which randomly selects an agent or action from the logs as the result, establishing a chance-level lower bound.

\noindent\(\bullet\) \textit{LLM-based Failure Attribution.}
This refers to the approaches proposed by Zhang et al.~\cite{who&when}.
Specifically, Zhang et al. employs three carefully designed prompting strategies to guide an LLM (GPT-4o) in generating failure attribution results from system logs: 1) \textit{All-at-once}: The LLM is prompted to directly identify the failure-responsible agent and action from the complete log. 2) \textit{Step-by-step}: The LLM processes the log step-by-step and is prompted to determine whether an error has occurred at the current step. This judging process terminates upon detecting the first mistake. 3) \textit{Binary-search}: The LLM initially processes the full log and is prompted to determine whether the error lies in the upper or lower half. This process is repeated recursively until the failure-responsible agent or action is identified.
Additionally, each prompting strategy is evaluated in two variants: with and without the inclusion of ground-truth answers in the prompt.
In total, there are six variants (3 strategies × 2 conditions) for the approach by Zhang et al..

\noindent\(\bullet\) \textit{\tech with Different Suspiciousness formula.} This refers to five variants of \tech by replacing its core scoring function $\mathbb{S}(\eta_j)$ with five traditional SBFL formulas: Ochiai~\cite{Ochiai}, Tarantula~\cite{tarantula}, Jaccard~\cite{Jaccard}, Dstar2~\cite{Dstar2}, and Kulczynski2~\cite{Kulczynski2}.
These variants are denoted as \tech-$Ochiai$, \tech-$Tarantula$, \tech-$Jaccard$, \tech-$Dstar2$, and \tech-$Kulczynski2$, respectively.




\noindent\textbf{Evaluation Metric.}
In line with related work~\cite{who&when}, we evaluate the accuracy of failure attribution at both the agent level and the action level. 
Specifically, the \textit{agent-level accuracy} measures the proportion of failure-responsible agents correctly identified by the attribution method.
The \textit{action-Level accuracy} calculates the percentage of decisive error actions accurately traced, providing a stricter evaluation than the agent-level metric.
For both these two accuracy, adopt a top-1 criterion~\cite{Dstar2}, i.e., a failure attribution is considered successful only if the ground-truth faulty triple is ranked first with no ties, for a strict evaluation.

\noindent\textbf{Implementation.}
We implement \tech in Python3.11, structuring the system into two core modules: a \textit{trajectory replay \& abstraction} module and a \textit{spectrum analysis} module. 
For the LLM used in \textit{trajectory replay \& abstraction} module, we consider the Qwen2.5-72B~\cite{qwen2.5} due to its open access, cost-efficiency, and local deployability—particularly crucial for replaying lengthy failure trajectories in handcrafted MASs.
For key parameters in \tech, unless otherwise mentioned, we set them as follows: (1) The number of repeated executions $k$ during trajectory replay is set to 20; (2) The decay factor $\lambda$ used in the suspiciousness scoring (Equations~\ref{eq:enhance} and~\ref{eq:decy}) is set to 0.9. 
A detailed sensitivity analysis of these and other parameters is provided in Section~\ref{sec:configuration}.
Moreover, since \tech requires re-executing the MASs to collect multiple execution trajectories for spectrum analysis, we follow the system information provided in the Who\&When benchmark to implement these MASs.
The two of the authors of this paper cross-check the implementations of these MASs to ensure their correctness.
Regarding the compared baselines, we directly adopt their open-source implementations and employ the default configurations of the original papers to ensure the accuracy of experimental repetition.
All the experiments were performed on a desktop equipped with an Intel$^{\circledR}$ Core$^{\text{TM}}$ i7-10700 CPU, 32GB RAM, running Ubuntu 22.04.

\subsection{RQ1: Effectiveness and Efficiency of \tech}

\begin{table}
	\centering
	\caption{Performance Comparison of Failure Attribution Accuracy (\%): SBFL-Based Methods (Including \tech and Its Variants) vs. LLM-Based Approaches (Unlabeled vs. Ground-Truth Labeled) and Random Approach at Agent and Action Levels on the Who\&When benchmark.}
	\label{tab:overall_performance}
	\resizebox{\textwidth}{!}{
	\begin{tabular}{llcccccc}
\toprule
		\multirow{2}{*}{\textbf{Method}} && \multicolumn{2}{c}{\textbf{Algorithmically Generated MASs}} & \multicolumn{2}{c}{\textbf{Hand-crafted MASs}} & \multicolumn{2}{c}{\textbf{Total}} \\
		\cmidrule(lr){3-4}
		\cmidrule(lr){5-6}
		\cmidrule(lr){7-8}
		&& \textbf{Agent-level} & \textbf{Action-level} & \textbf{Agent-level} & \textbf{Action-level} & \textbf{Agent-level} & \textbf{Action-level} \\
			\midrule
		- &Random &29.10 &19.06 &12.00 &4.16 &23.71 &14.36 \\
		\midrule
		\multirow{6}{*}{\makecell{LLM-\\Based}} 
		&All-at-Once  &51.12 &13.52 &53.44 &3.51 &51.85 &10.37 \\
		&Step-by-Step   &26.02 &15.31 &53.44 &8.77 &28.14 &13.25 \\
		&Binary Search    &30.11 &16.59 &36.21 &6.90 &32.03 &13.54 \\
		&All-at-Once ($\mathcal{G}$)   &54.33 &12.50 &55.17 &5.27 &54.59 &10.22  \\
		&Step-by-Step ($\mathcal{G}$)   &35.20 &\textbf{25.51} &34.48 &7.02 &34.97 &19.68 \\
		&Binary Search ($\mathcal{G}$)    &44.13 &23.98 &51.72 &6.90 &46.52 &18.60 \\
	\midrule
		  \multirow{6}{*}{\makecell{SBFL-\\Based}} 
&\tech-${Ochiai}$&50.79&19.84&\textbf{62.07}&25.86&54.35&21.74\\
		  &\tech-$Tarantula$&0.00&0.00&8.62&6.90&2.72&2.17\\
		  &\tech-$Jaccard$&50.79&19.84&58.62&22.41&53.26&20.65\\
		  &\tech-$Dstar2$&50.00&19.05&60.34&24.14&53.26&20.65\\
		  &\tech-$Kulczynski2$&50.79&19.84&\textbf{62.07}&24.14&54.35&21.20\\
	 \rowcolor{lightergray}&\tech&\textbf{55.56} &23.81 &\textbf{62.07} &\textbf{41.38} &\textbf{57.61} &\textbf{29.35} \\
		\bottomrule
	\end{tabular}
}
\end{table}
In this section, we evaluate the effectiveness and efficiency of \tech on the Who\&When benchmark.

\noindent\textbf{Effectiveness.} As shown in the last row of Table \ref{tab:overall_performance}, \tech demonstrates strong performance in failure attribution accuracy on the Who\&When benchmark. Evaluating on 184 failure execution logs, the method achieves 57.61\% (=106/184) accuracy at the agent level and 29.35\% (=54/184) accuracy at the action level. These results indicate that \tech can effectively identify faulty components in multi-agent systems across different granularity levels. The performance at the action level, while lower than at the agent level, reflects the increased difficulty of precisely localizing errors to specific actions within agent behaviors.

\noindent\textbf{Efficiency.} 
We measured the average time cost of \tech on the Who\&When benchmark. Overall, \tech takes approximately 105 minutes on average to complete a single failure attribution task, with execution times ranging from 38 minutes for simpler algorithm-generated MAS logs to 248 minutes for complex handcrafted MAS trajectories. 
In particular, each single failure attribution task of \tech includes two main stages: \textit{trajectory replay} \& \textit{abstraction} and \textit{spectrum analysis}. 
While the former stage is computationally intensive, the subsequent \textit{spectrum analysis} phase is highly efficient,  typically completing in under one minute once the abstracted execution trajectories are obtained. 
More specifically, the time cost of \textit{trajectory replay} \& \textit{abstraction} varies significantly depending on MAS complexity and the number of replay trials $k$ ($k=20$ in \tech). For the 126 algorithm-generated failure logs, the detailed average time cost is: 21 min for replay, 6 min for abstraction, and 11 min for clustering. In contrast, the 58 handcrafted logs require substantially more time: 136 min for replay, 103 min for abstraction, and 9 min for clustering. 


Overall, the results indicate that while \textit{spectrum analysis} scales efficiently, the \textit{trajectory replay} \& \textit{abstraction} stage remains the primary computational bottleneck, especially for handcrafted MAS trajectories that capture richer and more diverse behaviors. Nevertheless, the overall time cost remains acceptable, given that the process is fully automated and considering the inherent complexity of failure attribution in MAS.


\noindent\textbf{Generalizability.} \tech demonstrates consistent generalizability across different data sources within the Who\&When benchmark. On the algorithmically-generated MASs comprising 126 failure execution logs, the method achieves 55.56\% (=70/126) accuracy at the agent level and 23.81\% (=30/126) at the action level. In comparison, on the 58 handcrafted MAS failure logs, which typically exhibit greater complexity and longer execution sequences, \tech attains higher accuracy rates of 62.07\% (=36/58) at the agent level and 41.38\% (=24/58) at the action level.
Notably, the improvement in action-level accuracy on the more complex handcrafted MASs is particularly significant (from 23.81\% to 41.38\%). This performance pattern aligns with the characteristics of spectrum-based fault localization approaches, as the longer execution logs in handcrafted MASs provide richer spectral information for statistical analysis, thereby enhancing the method's ability to precisely localize faulty actions. The shorter logs typically found in algorithm-generated MASs (usually <10 steps) present a more challenging environment for action-level localization due to limited behavioral data.

\noindent\textbf{Answer to RQ1:} \tech demonstrates consistent effectiveness in failure attribution across different data sources within the Who\&When benchmark with tolerant time cost. And \tech achieves even higher accuracy when processing longer and more complex execution logs.


\subsection{RQ2: Comparison against Baselines}\label{sec:rq2}
In this section, we compare the failure attribution accuracy of \tech with that of 12 baseline methods on Who\&When benchmark.

\noindent\textbf{Effective Comparison} Table \ref{tab:overall_performance} also presents the failure attribution accuracy comparison between \tech and the other 12 baseline methods. The results demonstrate that \tech outperforms all baselines, achieving the highest accuracy at both the agent level (57.61\%) and action level (29.35\%) across the entire benchmark. More specifically, 

When compared to the random approach, \tech improves agent-level accuracy from 23.71\% to 57.61\%---a relative increase of 142.3\%—and raises action-level accuracy from 14.36\% to 29.35\%, representing a 104.4\% improvement. These results demonstrate that \tech significantly outperforms random attribution and confirms its capability to effectively capture meaningful fault patterns beyond random chance.

When compared to LLM-based approaches (using GPT-4o as the base model), \tech demonstrates substantial improvements across both evaluation levels. At the agent level, it outperforms the worst-performing LLM method (step-by-step) by 104.7\% (=29.47/28.14) and exceeds the best-performing LLM method (all-at-once with ground truth) by 5.5\% (=3.02/54.59). More notably, at the action level, \tech shows even more significant gains, surpassing the worst LLM method (all-at-once with ground truth) by 187.2\% (=19.13/10.22) and outperforming the best LLM method (step-by-step with ground truth) by 49.1\% (=9.67/19.68). Regardless of which LLM-based approach or evaluation level (agent or action) is considered, \tech demonstrates superior performance over all LLM-based methods on the Who\&When benchmark, establishing a clear and comprehensive advantage in failure attribution accuracy.

When compared to other variants of \tech using different suspiciousness formulas, \tech consistently demonstrates superior performance. At the agent level, it achieves a 6.0\% (=3.26/54.35) improvement over the best-performing variants (\tech-$Ochiai$ and \tech-$Kulczynski2$). More notably, at the action level, \tech shows even more substantial gains, outperforming the top variant (\tech-$Ochiai$) by 35.0\% (=7.61/21.74). These results demonstrate that our multi-dimensional metric formula achieves significantly better accuracy in identifying faulty components, particularly in the more challenging task of precise action-level localization, highlighting the effectiveness of our integrated approach combining both agent behavior characteristics and action frequency patterns for MAS failure attribution.

A deeper analysis of the results from \tech and its variants reveals several important patterns. First, \tech-$Tarantula$ demonstrates the lowest failure attribution accuracy among all variants, primarily due to its reliance on successful execution trajectories. When no successful executions exist (e.g., in fully failing test suites), this variant fails to produce meaningful results. Second, all other variants of \tech (excluding \tech-$Tarantula$) outperform LLM-based approaches at the action level while maintaining comparable performance at the agent level. This dual-level superiority confirms SBFL's fundamental advantage for failure attribution in MAS.
\noindent\textbf{Generalizability Comparison} 
Across different data sources within the Who\&When benchmark, significant performance differences emerge. When applied to the 58 complex failure logs from handcrafted MASs, both the random approach and LLM-based methods show notable degradation in action-level attribution accuracy compared to their performance on the 126 simpler logs from algorithm-generated MASs. This decline is attributed to the substantially higher complexity of handcrafted MAS logs, which typically contain longer and more intricate execution sequences (from 3 to 65 steps) versus the relatively straightforward algorithm-generated logs (<10 steps). 

By contrast, SBFL-based approaches—particularly \tech and its variants—demonstrate consistent performance across both data sources. Remarkably, \tech not only maintains but improves its accuracy on the more challenging handcrafted MAS logs, achieving 62.07\% agent-level and 41.38\% action-level accuracy. This represents a 371.8\% (=32.61/8.77) improvement over the best LLM-based approach (step-by-step with ground truth) on the same data at action-level. This robustness highlights the advantage of spectrum-based analysis in handling complex MAS environments where LLM-based methods struggle.

 Regarding performance across different granularities of fault attribution, random approach shows comparable action-level performance comparable to some LLM-based methods, yet fails to effectively capture agent-level fault patterns. Most LLM-based approaches exhibit a significant trade-off: they either achieve high agent-level accuracy at the expense of action-level precision, or improve action-level detection while suffering degradation in agent-level performance. In contrast, SBFL-based approaches, particularly \tech and its variants, maintain strong and consistent performance across both granularities simultaneously. This consistent superiority highlights a crucial advantage of spectrum-based methods: unlike LLM-based approaches that struggle to balance dual-level accuracy, SBFL-based methods achieve robust performance at both agent and action levels, demonstrating their effectiveness for comprehensive fault localization in complex MAS environments.

 \noindent\textbf{Answer to RQ2:} Compared to all baseline techniques, \tech demonstrates superior performance, outperforming the random approach, all LLM-based methods, and even its own variants using different suspiciousness formulas. Furthermore, it shows significantly better generalization capabilities across diverse data sources and different levels of fault granularity.


\subsection{RQ3: Configurations of \tech}~\label{sec:configuration}
In this section, we evaluate the influence of key parameters on the performance of \tech and conduct an ablation study to assess the contribution of different components in the suspiciousness scoring formula. To more intuitively demonstrate the impact of different parameters and components---particularly where accuracy differences may appear small yet practically meaningful—we report the number of successful failure attributions (i.e., the count of logs where the ground---truth faulty triple is uniquely ranked first) rather than the accuracy rate.

\begin{figure}[!tb]
    \centering
    \begin{minipage}{0.45\textwidth}
        \centering
        \begin{tikzpicture}
        \begin{axis}[
            width=\textwidth,
            height=4cm,
            ymin=95, ymax=110,
            xmin=2, xmax=21,
            xtick={5,10,15,20},
            xlabel={Number of Replayed Trajectory $k$},
            ylabel={Successful Attribution},
            xlabel style={font=\fontsize{8}{7}\selectfont\sffamily}, 
    ylabel style={
            at={(axis description cs:-0.15,0.5)},
            yshift = -10pt,
            font=\fontsize{8}{7}\selectfont\sffamily},
            ymajorgrids=true,
            grid style=dashed,
            legend style={at={(0.5,1.05)}, anchor=south, legend columns=2, font=\scriptsize\selectfont\sffamily},
            legend image post style={draw=black},
            tick label style={font=\scriptsize}
        ]
        \addplot+[ybar stacked, bar width=15pt, fill=none, draw=none, mark=none, forget plot] 
            coordinates {(5,97) (10,98) (15,102) (20,106)};
        \addplot+[ybar stacked, bar width=15pt, mark=none, draw=none, fill=myblue] 
            coordinates {(5,5) (10,7) (15,6) (20,0)};
        \addlegendentry{Min--Max Range}
        \addplot+[thick, color=black, mark=*, mark options={draw=black, fill = blue}] coordinates {
            (5,99.6) (10,101.4) (15,104.6) (20,106)
        };
        \addlegendentry{Average}
        \end{axis}
        \end{tikzpicture}
        \subcaption{Agent-level}
    \end{minipage}
    \hfill
    \begin{minipage}{0.45\textwidth}
        \centering
        \begin{tikzpicture}
        \begin{axis}[
            width=\textwidth,
            height=4cm,
            ymin=42, ymax=56,
            xmin=2, xmax=21,
            xtick={5,10,15,20},
            xlabel={Number of Replayed Trajectory $k$},
            ylabel={Successful Attribution},
            xlabel style={font=\fontsize{8}{7}\selectfont\sffamily}, 
    ylabel style={
    at={(axis description cs:-0.15,0.5)},
    yshift = -10pt,
    font=\fontsize{8}{7}\selectfont\sffamily},
            ymajorgrids=true,
            grid style=dashed,
            legend style={at={(0.5,1.05)}, anchor=south, legend columns=2, font=\scriptsize\selectfont\sffamily},
            legend image post style={draw=black},
            tick label style={font=\scriptsize}
        ]
        \addplot+[ybar stacked, bar width=15pt, fill=none, draw=none, mark=none, forget plot] 
            coordinates {(5,44) (10,46) (15,49) (20,54)};
        \addplot+[ybar stacked, bar width=15pt, mark=none, fill=mygreen, draw=none] 
            coordinates {(5,5) (10,3) (15,5) (20,0)};
        \addlegendentry{Min--Max Range}
        \addplot+[thick, color=black, mark=*, mark options={draw=black, fill = red}] coordinates {
            (5,46) (10,47.6) (15,50.8) (20,54)
        };
        \addlegendentry{Average}
        \end{axis}
        \end{tikzpicture}
        \subcaption{Action-level}
    \end{minipage}
    \caption{Effect of The Number of Replayed Trajectory $k$ on \tech's Performance.}
    \label{fig:trace_length_split}
\end{figure}
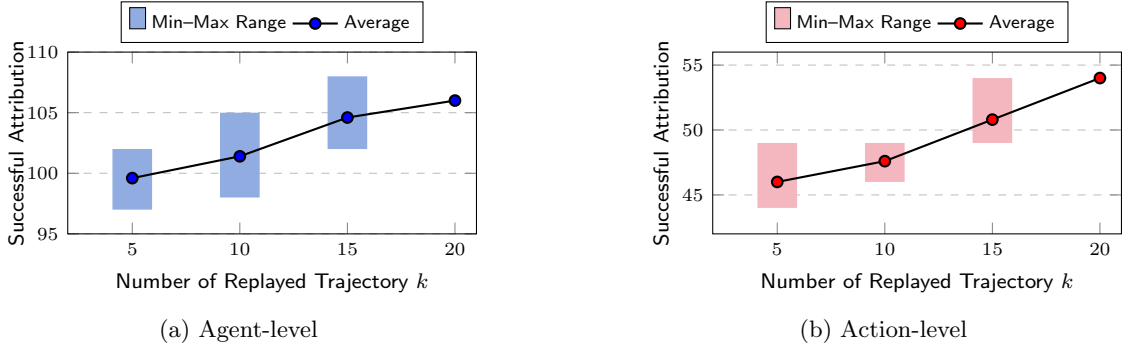

\noindent\textbf{Effect of The Number of Replayed Trajectory $k$.} We evaluate the impact of the number of repeated executions $k$ during the trajectory replay phase by testing \tech's performance with different values of this parameter. Specifically, we assess \tech on the Who\&When benchmark by randomly sampling 5, 10, or 15 trajectories per task query, with each configuration repeated 5 times to ensure statistical reliability. Figure~\ref{fig:trace_length_split} shows that the average number of correctly attributed failures increases with additional execution trajectories: At the agent level: averages of 99.6 ($k$=5), 101.4 ($k$=10), and 104.6 ($k$=15), with respective ranges of [97-102], [98-105], and [102-108].
At the action level: averages of 46.0 ($k$=5), 47.6 ($k$=10), and 50.8 ($k$=15), with respective ranges of [44-49], [46-49], and [49-54]. These results indicate that \tech maintains robustness with moderate reductions in trajectory quantity, though both quantity and representativeness of trajectories impact effectiveness. Performance improves with larger $k$ values, as more diverse execution trajectories enhance spectrum analysis—particularly for action-level attribution where richer behavioral patterns enable better statistical correlation.

\begin{wrapfigure}{r}{0.43\textwidth}
    \begin{tikzpicture}
        \begin{axis}[%
            width=0.32\textwidth,  
            height=0.22\textwidth, 
            xlabel={Decay Factor $\lambda$},
            ylabel={Agent-level},
            xmin=0.65, xmax=1.00,
            ymin=102, ymax=107,
            xtick={0.65,0.70,0.75,0.80,0.85,0.90,0.95,1.00},
            ytick={103,104,105,106},
            ylabel style={at={(axis description cs:-0.25,0.55)},yshift=-18pt,font=\fontsize{8}{7}\selectfont\sffamily},
            xlabel style={font=\fontsize{8}{7}\selectfont\sffamily},
            axis y line*=left,
            axis x line=bottom,
            xmajorgrids=true,  
grid style={dashed, lightgray!80!white, line width=0.6pt},
            scale only axis,
            legend style={font=\scriptsize\selectfont\tiny, at={(0.05,0.95)}, anchor=north west},
            tick label style={font=\scriptsize}
        ]
        \addplot[ultra thick, color=myblue,mark=*, mark options={fill = blue}] coordinates {
            (0.65,104) (0.70,104) (0.75,104) (0.80,104) (0.85,105) (0.90,106) (0.95,106) (1.00,103)
        };
        \addlegendentry{Agent-level}

        \addlegendimage{ultra thick,color=mygreen,mark=square*,mark options={fill = red}}
        \addlegendentry{Action-level}
        \end{axis}

        \begin{axis}[%
            width=0.32\textwidth,
            height=0.22\textwidth,
            at={(current axis.south west)}, 
            anchor=south west,
            axis y line*=right,
            axis x line=none,
            xmin=0.65, xmax=1.00,
            ymin=45, ymax=55, 
            ytick={46,48,50,52,54},
            ylabel={Action-level},
            ylabel style={at={(axis description cs:1.15,0.3)},anchor=west,color=black,font=\fontsize{8}{7}\selectfont\sffamily},
            tick label style={font=\scriptsize},
            yticklabel style={color=black},
            scale only axis
        ]
        \addplot[ultra thick, color=mygreen,mark=square*,mark options={fill = red}] coordinates {
            (0.65,47) (0.70,49) (0.75,49) (0.80,50) (0.85,53) (0.90,54) (0.95,54) (1.00,46)
        };
        \end{axis}
    \end{tikzpicture}
    \caption{Effect of The Decay Factor $\lambda$.}
    \label{fig:lambda_ablation}
\end{wrapfigure}
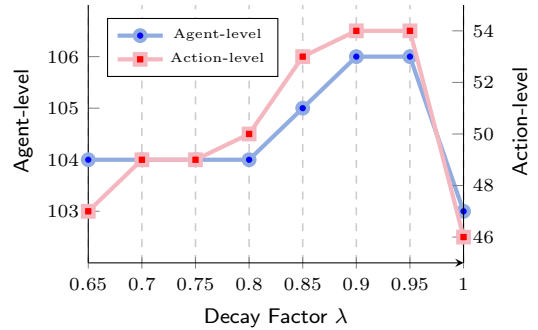

\noindent\textbf{Effect of The Decay Factor $\lambda$.} We evaluate the impact of the \textit{decay factor} ($\lambda$), which plays an important role in the suspiciousness calculation formula, on \tech's performance at both agent and action levels, as shown in Figure~\ref{fig:lambda_ablation}.
As $\lambda$ increases from 0.65 to 0.95, the number of successful agent-level attributions steadily rises from 104 to 106, while action-level attributions increase from 47 to 54, reaching peak performance around $\lambda = 0.90\text{--}0.95$. When $\lambda$ reaches 1.00, both metrics drop (agent-level: 103, action-level: 46), indicating that completely ignoring frequency decay (i.e., treating all occurrences equally) reduces attribution performance.
Overall, these results demonstrate that \tech's effectiveness is sensitive to the $\lambda$ parameter, with moderate values ($\lambda \approx 0.9$) providing the optimal balance for capturing both agent-level and action-level fault patterns through appropriate frequency weighting.

\begin{wraptable}{r}{0.4\textwidth}
	\centering
	\caption{Performance of \tech Variants with Different Parameter Combinations.}
\label{tab:ablation_parameters}
	\resizebox{\linewidth}{!}{
		\begin{tabular}{lccccc}
			\toprule
			\multirow{2}{*}{\textbf{Tools}} & \multirow{2}{*}{\textbf{$\lambda$}} &  \multirow{2}{*}{\textbf{$\gamma$}} &\multirow{2}{*}{\textbf{$\beta$}} & \multicolumn{2}{c}{\textbf{Who\&When}}\\
            \cmidrule(lr){5-6}
            &&&&\textbf{Agent-level} & \textbf{Action-level}\\
			\midrule
			\tech-$K$ & \ding{55} & \ding{55} & \ding{55} & 100&39 \\
			\midrule
			\tech-${O\beta}$  & \ding{51}& \ding{51} & \ding{55} & 106&51 \\
			\tech-${O\gamma}$  & \ding{51}& \ding{55} & \ding{51} &103&51 \\
            \tech-${O\lambda}$  & \ding{55} & \ding{51} & \ding{51} &104&43 \\
			\midrule
			\rowcolor{lightergray} {\tech}  & \ding{51} & \ding{51} &\ding{51}& 106&54 \\
			\bottomrule
		\end{tabular}
		}
	\label{tab:component}
\end{wraptable}
\noindent\textbf{Effect of Components in the Suspiciousness Formula.} We perform an ablation study to quantify the contribution of three key components in the suspiciousness formula (see Equation \ref{final_eq}) specifically designed for MAS: the \textit{decay factor} ($\lambda$), the \textit{action coverage ratio} ($\gamma$), and the \textit{action frequency proportion} ($\beta$), introduced in Section~\ref{sec:fa4MAS}. We create 4 different variants of \tech: (1) \tech-$K$, which uses the base Kulczynski2 formula only; (2) \tech-${O\beta}$, which removes the \textit{action coverage ratio} $\gamma$; (3) \tech-${O\gamma}$, which removes the \textit{action frequency proportion} $\beta$; and (4) \tech-${O\lambda}$, which specially recovers the $\lambda$-decay in the Kulczynski2 formula while removing the \textit{local enhancement factor} ($\alpha_{\tau}$). Table~\ref{tab:ablation_parameters} reports the agent-level and action-level performance of \tech\ variants with different combinations of these components.  

\tech significantly outperforms the base formula, improving agent-level results from 100 to 106 and action-level results from 39 to 54. When examining the three ablated variants, we observe that removing any component from the formula decreases performance, with action-level results dropping to 43–51. The removal of $\lambda$ leads to the largest decline, highlighting its critical role. At the agent level, the results are more stable, ranging from 103 to 106 across variants. Overall, all three components contribute to improved performance, and their combination yields the best results.

\noindent\textbf{Answer to RQ3:} (1) Reducing the number of replayed trajectory $k$ moderately (e.g., from 20 to 15) has limited effect to \tech, but further reduction decreases action-level performance. (2) The decay factor ($\lambda$) plays a critical role: moderate values (0.9-0.95) yield the highest agent-level and action-level results, where extreme values lead to reduced performance. (3) All components of the curated suspiciousness formula contribute positively.

\section{Discussion}
\subsection{Impact of Log Complexity on Failure Attribution Performance}

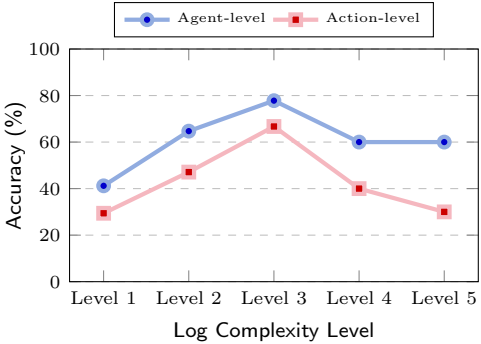
\begin{wrapfigure}{r}{0.43\textwidth}
    \begin{tikzpicture}
    \begin{axis}[
        width=0.45\textwidth,  
        height=0.3\textwidth, 
        xlabel={Log Complexity Level},
        ylabel={Accuracy (\%)},
        ylabel style={at={(axis description cs:-0.2,0.5)},yshift=-18pt,font=\fontsize{8}{7}\selectfont\sffamily},
        xlabel style={font=\fontsize{8}{7}\selectfont\sffamily},
        xtick={1,2,3,4,5},
        xticklabels={Level 1, Level 2, Level 3, Level 4,Level 5},
        ymin=0, ymax=100,
        legend style={at={(0.5,1.05)}, anchor=south, legend columns=2, font=\scriptsize\selectfont\tiny},
        tick label style={font=\scriptsize},
        ymajorgrids=true, grid style=dashed,
    ]

    \addplot+[ultra thick, color=myblue, mark=*] coordinates {(1,41.2) (2,64.7) (3,77.8) (4,60) (5,60)};
    \addlegendentry{Agent-level}
    \addplot+[ultra thick, color=mygreen, mark=square*] coordinates {(1,29.4) (2,47.1) (3,66.7) (4,40) (5,30)};
    \addlegendentry{Action-level}
    \end{axis}
    \end{tikzpicture}
    \caption{Effect of log complexity on \tech’s agent-level and action-level accuracy.}
    \label{fig:log_length}
\end{wrapfigure}

From Table \ref{tab:overall_performance}, we observe that \tech achieves higher performance on handcrafted logs compared to those generated by algorithmic MASs. This difference is particularly pronounced at the action level. This performance gap is largely attributed to the constrained maximum step count (limited to 10) in algorithm-generated MASs, where spectrum-based approaches are generally less effective in highly simplified scenarios. To further investigate the relationship between task complexity and performance, we categorized the handcrafted logs into five distinct complexity levels (Level 1–Level 5) based on step count: Level 1 (0–11 steps), Level 2 (12–23 steps), Level 3 (24–37 steps), Level 4 (38–51 steps), and Level 5 (52–65 steps). The corresponding agent-level and action-level accuracies for each complexity level are provided in Figure~\ref{fig:log_length}.

Our analysis reveals that \tech exhibits a notable sensitivity to log complexity. Both agent-level and action-level accuracies are lowest for the most simplistic logs (Level 1), owing to limited contextual information available for effective failure attribution. Performance peaks at moderate complexity (Levels 2-3), where there is sufficient context to identify failure-responsible agents and actions without being overwhelmed by excessive noise. With further increases in complexity (Levels 4–5), accuracy metrics decline again, reflecting the difficulty of isolating decisive steps in longer, more intricate trajectories. This pattern suggests that \tech is most effective in scenarios with balanced task complexity, while extremely simple or highly complex MAS settings remain challenging. These trends are consistent with findings from prior SBFL studies on program complexity~\cite{SBFL_study}, where extremely simple or highly complex programs also tend to yield lower localization accuracy.

To address these limitations, future work could explore techniques such as hierarchical context modeling or selective trajectories pruning to enhance performance across the full spectrum of log complexities.

\subsection{Threads to Validity}
The main threat to \textbf{internal} validity lies in the correctness of the implementation of \tech, the compared approaches, and experimental scripts.
To reduce this threat, we adopt the open-source implementations of the compared approaches and build our approach on state-of-the-art libraries, and carefully check the source code of \tech and the experimental scripts.
The main threat to \textbf{external} validity lies in the selection of subjects in our study. 
To mitigate this threats, we perform on the recently proposed Who\&When benchmark.
This benchmark provides diverse failed execution trajectories across various MAS types, offering comprehensive coverage of realistic scenarios.
Furthermore, all failed execution trajectories in this benchmark are manually annotated by human experts through a multi-stage consensus procedure, ensuring annotation accuracy.
The main threat to \textit{construct} lies in the metrics used in our experiments and the parameters in \tech.
To reduce the threat from metrics employed, we evaluate the accuracy of failure attribution at both the agent level and the action level.
Both of these two-level accuracies are widely employed in related work~\cite{who&when, agenttracer}.
To reduce the threat from the parameters in \tech, we present the detailed parameter settings in Section~\ref{sec:exp_setup} and investigate the impact of these parameters in Section~\ref{sec:configuration}.


\section{Related Work}

\noindent\textbf{Failure Analysis in MASs.}
In light of growing concerns regarding the reliability of MASs, a series of papers~\cite{mast, trail,who&when, agenttracer}came out recently that analyzes the failures in MASs.
MAST~\cite{mast} is the first to comprehensively characterize failure executions in MASs, developing a failure taxonomy comprising 14 failure modes across system design, agent coordination, and task verification. 
Following MAST, TRAIL~\cite{trail} introduces a more fine-grained taxonomy of failures, encompassing reasoning, planning, coordination, and system execution in MASs.
AGDebugger~\cite{microsoft_interactive_debugging_and_steering_MAS} introduces an interactive tool enabling developers to debug and steer agent teams by inspecting and editing message histories.
More recently, Zhang at al.~\cite{who&when} formalize the automated failure attribution problem for MASs and propose leveraging the LLM-as-a-judge paradigm to attribute failures from system logs.
Additionally, AgentTracer~\cite{agenttracer}, a concurrent work with \tech, further fine-tunes the LLM for failure attribution in MASs.
In contrast to the prior techniques that directly employ the LLMs for failure attribution, \tech draws inspiration from traditional spectrum-based fault localization, applying spectrum analysis to achieve more precise and effective failure attribution.

\noindent\textbf{Testing of MASs.}
Dozens of testing techniques have been proposed for MASs, which can be roughly classified into two families, namely benchmarks and red-teaming.
The benchmarks~\cite{agentbench,multiagentbench,agentsnet, swe-bench, gaia} aim to develop challenging tasks reflecting real-world MAS applications to evaluate system effectiveness.
Representative examples include AgentBench~\cite{agentbench} for general tasks, SWE-Bench~\cite{swe-bench} for software engineering, and GAIA~\cite{gaia} for assistants.
In contrast, the red-teaming focuses on adversarially probing these systems to uncover vulnerabilities and potential misuses through jailbreaking~\cite{jailbreaking_1,jailbreaking_2,jailbreaking_3} and prompt injection~\cite{prompt_injection_1,prompt_injection_2}. 
Despite the advancement of these testing approaches, failure attribution remains largely unexplored, hindering systematic optimization and improvement. 
This work addresses the overlooked failure attribution problem in MASs by proposing a novel spectrum-based approach \tech.

\noindent\textbf{Fault Localization in Traditional Software.}
In the field of traditional software engineering, various techniques have been proposed for localizing faults, such as spectrum-based~\cite{sbfl_survey, Ochiai, tarantula, Jaccard, Dstar2, Kulczynski2}, mutation-based~\cite{mutate_fl_1,mutate_fl_2,mutate_fl_3}, program-slicing-based~\cite{slicing_fl_1,slicing_fl_2,slicing_fl_3} and learning-based~\cite{learning_fl_1, learning_fl_2, learning_fl_3, learning_fl_4}.
Among these, spectrum-based fault localization (SBFL) is one the mostly widely stuided one in the literature, due to its lightweight and scalability~\cite{sbfl_popular}.
\tech extends traditional SBFL to emerging MASs by proposing a novel suspiciousness formula tailored specifically for MASs.
\section{Conclusion}
In this paper, we propose \tech, the first spectrum-based failure attribution approach for multi-agent systems (MASs).
\tech identify the root cause of a specific failed trajectory by performing spectrum analysis on multiple trajectories collected through repeated execution of the corresponding task.
In particular, \tech implements a novel suspiciousness formula that captures both agent activation patterns and action activation patterns in the system execution trajectories.
Experimental results on the Who\&When benchmark demonstrate the effectiveness of \tech, where \tech achieves the best performance compared to in total 12 baselines.

\bibliographystyle{plain}
\bibliography{main}
\end{document}